\documentclass{article}
\usepackage[utf8]{inputenc}
\usepackage[text={8.5in,11in},margin=1.in,includefoot]{geometry}
\usepackage{graphicx}
\usepackage{float}
\usepackage{amsmath}
\usepackage{hyperref}
\usepackage{authblk}
\usepackage[hang,flushmargin]{footmisc}

\title{JILA SrI Optical Lattice Clock With Uncertainty of 2.0$\times 10^{-18}$}

\author[1]{Tobias Bothwell$^*$}
\author[1]{Dhruv Kedar$^*$}
\author[1]{Eric Oelker}
\author[1]{John M. Robinson} 
\author[1]{\\ Sarah L. Bromley$^\dagger$}
\author[1, 2]{Weston L. Tew}
\author[1]{Jun Ye}
\author[1]{Colin J. Kennedy}

\affil[1]{\footnotesize JILA, National Institute of Standards and Technology and University of Colorado, Boulder, CO 80309, USA}
\affil[2]{\footnotesize National Institute of Standards and Technology, Gaithersburg, Maryland 20899, USA}

\date{\small \today \vspace{-8ex}}

\begin{document}

\maketitle

{\let\thefootnote\relax\footnote{{$^*$ Equal Contributions \\ $^\dagger$ Present Address:  Department of Physics, Durham University, Durham, UK \\ Corresponding author: tobias.bothwell@colorado.edu}}}

\begin{abstract}
We report on an improved systematic evaluation of the JILA SrI optical lattice clock, achieving a nearly identical systematic uncertainty compared to the previous strontium accuracy record set by the JILA SrII optical lattice clock (OLC) at $2.1\times10^{-18}$. This improves upon the previous evaluation of the JILA SrI optical lattice clock in 2013, and we achieve a more than twenty-fold reduction in systematic uncertainty to $2.0\times10^{-18}$. A seven-fold improvement in clock stability, reaching $4.8\times10^{-17}/\sqrt{\tau}$ for an averaging time $\tau$ in seconds, allows the clock to average to its systematic uncertainty in under $10$ minutes. We improve the systematic uncertainty budget in several important ways. This includes a novel scheme for taming blackbody radiation-induced frequency shifts through active stabilization and characterization of the thermal environment, inclusion of higher-order terms in the lattice light shift, and updated atomic coefficients. Along with careful control of other systematic effects, we achieve low temporal drift of systematic offsets and high uptime of the clock. We additionally present an improved evaluation of the second order Zeeman coefficient that is applicable to all Sr optical lattice clocks. These improvements in performance have enabled several important studies including frequency ratio measurements through the Boulder Area Clock Optical Network (BACON), a high precision comparison with the JILA 3D lattice clock, a demonstration of a new all-optical time scale combining SrI and a cryogenic silicon cavity, and a high sensitivity search for ultralight scalar dark matter.
\end{abstract}

\section{Introduction}

Atomic clocks based on optical transitions have realized unprecedented levels of stability, reproducibility, and accuracy \cite{bloom2014,nicholson2015systematic,ushijima2015cryogenic,mcgrew2018atomic,sanner2019optical,brewski}. This has enabled their use in a variety of applications ranging from the proposed redefinition of the SI second \cite{LodewyckSIsecond,Falke, BlattSI}, to searches for variations of fundamental constants \cite{derevianko2014hunting,wcislo2017experimental,roberts2017search,Kennedy}, and increased capabilities for positioning, navigation, and timing applications \cite{deschenes2016synchronization}.

However, in order to fully realize the potential for these applications, controlling the temporal drift of systematic offsets poses a potential barrier to control of the system at and below the $10^{-18}$ level. While it is possible to evaluate the systematic uncertainty of optical clocks at this level, active control of all systematics is now urgently needed~\cite{Grebing,IRHill} to realize a robust optical frequency reference that maintains this level of uncertainty over long time periods. A systematic shift can be measured to an extremely high precision but if it varies significantly over time the clock frequency will appear to drift unless real-time frequency corrections are applied. An alternative approach is to instead control the atom environment to the level where the temporal variations of systematic effects are well within the clock's uncertainty budget and frequency post-corrections need not be applied. This is the guiding principle taken in upgrading the JILA SrI optical lattice clock. Here, we demonstrate clock performance where fluctuations in systematic offsets are routinely bounded below $4 \times 10^{-19}$ for times up to $10^4$ seconds.

In implementing these upgrades the systematic uncertainty of the SrI clock, last evaluated at 5.3$\times 10^{-17}$ in 2013 \cite{bloom2014}, has been improved by more than a factor of 20 to 2.0$\times 10^{-18}$. This surpasses the uncertainty record for a strontium optical lattice clock and places this clock among the most accurate clocks in the world \cite{nicholson2015systematic, mcgrew2018atomic, brewski}. The active control of clock operating conditions, including the thermal environment, allows the clock to run in a robust manner without needing real-time frequency corrections to average to the level of its accuracy. In this paper, we detail the control of each major systematic effect in JILA SrI.

\section{Experimental Methods}

The experimental sequence begins with a collimated atomic beam of Sr generated from a effusive oven. Atoms of $^{87}$Sr in the beam undergo 2D transverse cooling and Zeeman slowing using the broadband 461\hspace{0.5mm}nm $^1S_0$\hspace{0.5mm}-\hspace{0.5mm}$^1P_1$ transition ($\Gamma = 2\pi\times32$\hspace{0.5mm}MHz). The slowed atoms are then loaded into a magneto-optical trap (MOT) operating on the same transition, and cooled to $\textrm{mK}$-level temperatures. After 100\hspace{0.5mm}ms of loading, the atoms are transferred to a second narrow-linewidth ($\Gamma = 2\pi\times7.5$\hspace{0.5mm}kHz) MOT operating on the dipole-forbidden $^1S_0$\hspace{0.5mm}-\hspace{0.5mm}$^3P_1$ line at 689\hspace{0.5mm}nm, cooling the atoms to a temperature of $3\hspace{0.5mm}\mu\textrm{K}$. The sample is then transferred to a one-dimensional, 813\hspace{0.5mm}nm red-detuned optical lattice, which exhibits minimal differential polarizability for the two clock states, and is optically pumped into one of the $|F=\frac{9}{2}, m_F=\pm\frac{9}{2}\rangle$ stretched states of the $^1S_0$ manifold. The lattice is initially loaded at a depth of $180\hspace{0.5mm}\text{E}_\text{r}$ to maximize the number of captured atoms and is then adiabatically ramped down to a nominal operating depth of $45\hspace{0.5mm}\text{E}_\text{r}$ (where $\text{E}_\text{r}$ is the recoil energy associated with the 813 nm trapping light) to minimize the effects of systematic shifts associated with the trapping light.

With state preparation complete, clock spectroscopy is performed on the $\sim 1$~mHz natural linewidth, 698\hspace{0.5mm}nm $^1S_0$ - $^3P_0$ transition using a narrow-linewidth ultrastable laser.  A bias magnetic field of 570 mG splits the $|\frac{9}{2},\pm\frac{9}{2}\rangle$ states by 556 Hz, and the clock transition is interrogated with a 600 ms long $\pi$-pulse with polarization collinear with the quantization axis. After interrogation with the clock laser, the resulting excitation fraction is measured by first detecting the ground state population via fluorescence on the 461 nm transition, and then detecting the excited state population by repumping atoms back to the ground state and again collecting fluorescence on the 461 nm transition.  The resulting difference in excitation fraction specifies the detuning of the spectroscopy laser from resonance. By alternately probing the two stretched states, we can reject fluctuations in the first-order Zeeman shift arising from magnetic field noise, further details of which are offered in \cite{Oelker}.

\begin{figure}[!htb]
\centering
\includegraphics[scale=0.75]{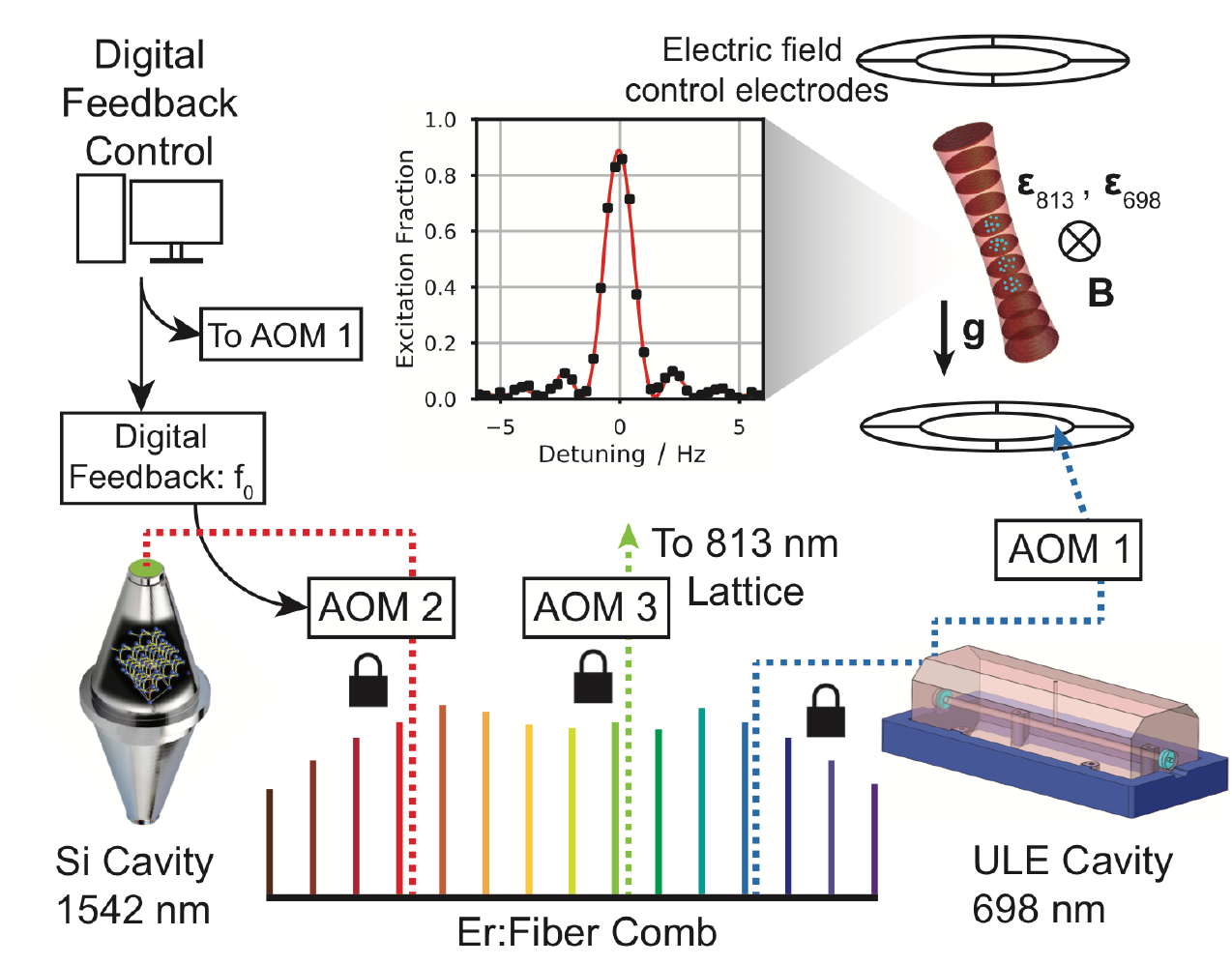}
\caption{\textbf{Schematic view of the SrI clock}. Ultrastable laser light is generated at 1542 nm by referencing a diode laser to a crystalline silicon optical cavity operating at 124 K (red, dotted line). The stability of this laser is then transferred via an Er:fiber comb to an external-cavity diode laser pre-stabilized by a 40 cm ULE cavity operating at 698 nm (blue, dotted line). An acousto-optic modulator (AOM 1) is then used to steer the cavity light into resonance with the Sr clock transition. The excitation fraction after probing the clock transition is detected by collecting fluorescence from both ground and excited state atoms. A frequency step applied to AOM 1 produces an error signal for locking by alternately probing both sides of the $\vert \pm 9/2 \rangle$ stretched state transitions. Frequency corrections to the average of the $\vert \pm 9/2 \rangle$ frequencies are applied to AOM 2 such that the cavity-stabilized light is steered onto the transition frequency of the Sr atom. In addition, frequency corrections to the difference of the $\vert \pm 9/2 \rangle$ frequencies are applied to the AOM 1 frequency. An in-plane magnetic field, $\mathbf{B}$, providing a quantization axis for the atoms, is aligned to be collinear with both the 1D optical lattice polarization, $\mathbf{\epsilon}_{813}$, and the clock laser polarization, $\mathbf{\epsilon}_{698}$. Out-of-vacuum quadrant ring electrodes generate a DC electric field to cancel the ambient field at the position of the atoms. Finally, a phase lock of the 813 nm trapping laser to the Er:fiber comb stabilizes the frequency of the trapping light (green, dotted line). The trapping light is delivered to the atoms through a high power optical fiber and is intensity stabilized by actuating the RF power on AOM 3.}
\label{fig:Schematic} 
\end{figure}

As depicted in Fig.~\ref{fig:Schematic}, the ultrastable laser used to probe the clock transition consists of a pre-stabilized 698\hspace{0.5mm}nm laser locked to a commercial Er:fiber comb which is phase-stabilized to a master 124 K silicon cavity. A frequency step applied to an acousto-optic modulator (AOM 1 in Fig.~\ref{fig:Schematic}) alternately probes either side of the $\vert \frac{9}{2},\pm\frac{9}{2}\rangle$ transitions and generates an error signal. A digital servo filter (PI$^2$D) is applied to this error signal to apply frequency feedback to AOM 2 in Fig.~\ref{fig:Schematic}. This loop configuration has the effect of stabilizing both the light after the AOM 2 and the frequency of each comb tooth to the spectroscopic precision and accuracy of clock operation. Under typical operating conditions (i.e. sample preparation time of 570 ms, interrogation time of 600 ms, and an atom number of $N = 1000$) the clock achieves a stability of $4.8\times10^{-17}/\sqrt{\tau}$ \cite{Oelker}.

\section{Systematic Evaluation}

Accurate determination of the unperturbed $^{87}$Sr clock transition frequency requires characterization of all systematic effects which produce energy shifts between the $^{1}\text{S}_{0}$ ground state and the $^{3}\text{P}_{0}$ metastable excited state. These effects range from the interaction of a single atom with an external field to two-particle collisions and many-body effects. For each shift, the perturbing effect is either directly measured - as in the case of the thermal electric field produced by room temperature radiation - or the effect is inferred by modulation of the applied field and the corresponding measurement of a frequency shift - as in the case of the lattice light shift. Subsequently, the systematic shifts relative to the unperturbed atomic transition frequency can be extrapolated to daily operating conditions using well-characterized theoretical models of each effect.

\begin{figure}[!htb]
\centering
\includegraphics[scale=0.315]{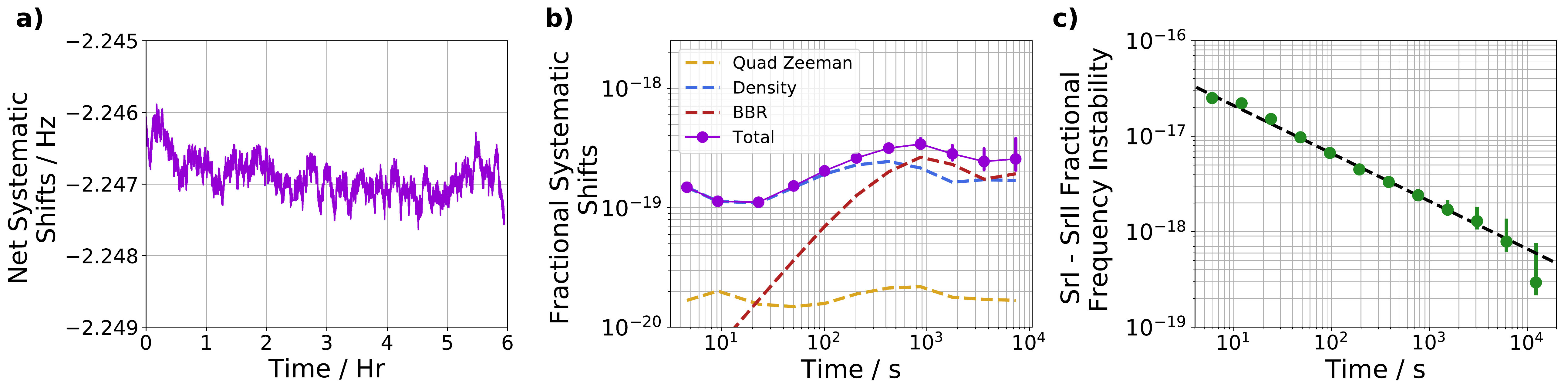}
\caption{\textbf{Systematic shifts. (a)} Plot of the time record of the systematic shifts. Changes in atom number, ambient temperature, or magnetic field all result in corrections to the clock frequency, and their total magnitude is shown over a six hour data campaign. The clock achieves 98.9$\%$ uptime over the course of this single comparison day and slight gaps in the data indicate brief periods where the laser is not locked to the atoms. \textbf{(b)} The same data is plotted as a fractional instability normalized to the Sr clock frequency. The individual contributions of density shift (blue), BBR (red) and second order Zeeman shift (yellow) are shown as the dashed curves. For operation times up to $10^4$ seconds, fluctuations in systematic offsets are bounded below $4 \times 10^{-19}$. \textbf{(c)} Nonsynchronous comparison with the JILA 3D optical lattice clock demonstrates that the beat between the two clocks averages below the quoted total systematic uncertainty. All error bars are derived from a white noise model and the black line is a white noise $\tau^{-1/2}$ fit to the single clock instability.}
\label{fig:AtomCorr}
\end{figure}

For shifts evaluated using the lock-in technique, the record-low clock instability demonstrated in Ref.~\cite{Oelker} allows the rapid determination of these systematic offsets - without the need to apply a large lever arm - at the $1\times10^{-18}$ level in less than one hour. However, the capability to rapidly evaluate shifts does not preclude the need for applying real-time frequency corrections to compensate for non-stationary systematics. A systematic offset can have temporal variations if its source (e.g. atom number) or calibration (e.g. fluorescence to atom number conversion) fluctuates throughout the day. In the following subsections, we show how, by active stabilization of the thermal environment and careful control of the operating parameters, drifting systematic offsets can remain below $4\times10^{-19}$ fractional instability over six hours of operation (Figs.~\ref{fig:AtomCorr}a, b). Consequently, a frequency comparison of the SrI clock against a stable reference (the JILA 3D clock \cite{campbell2017fermi}), Fig.~\ref{fig:AtomCorr}c, can average well into the $10^{-19}$ decade without systematic effects impacting the clock stability.

\subsection{Blackbody Radiation}\label{section:BBR}

\begin{figure}[h]
\centering
\includegraphics[scale=0.33]{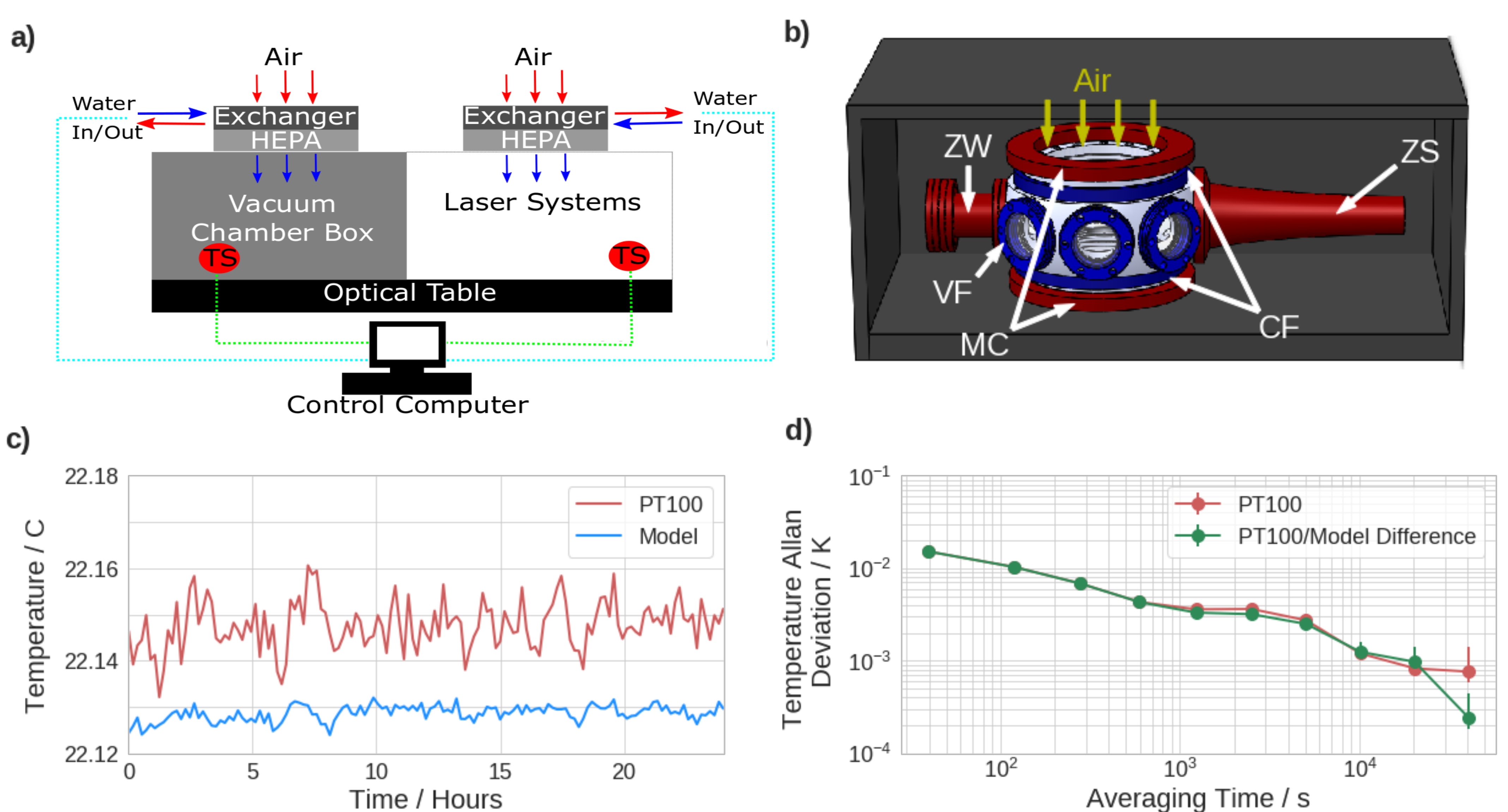}
\caption{\textbf{Active Temperature Control.} \textbf{(a)} The optical table is divided into independently controlled sections (grey and white shaded regions) and isolated from the room temperature by a laser curtain. Around the vacuum chamber is an additional black box. Temperature sensors (TS) monitor each table half, allowing feedback (dotted lines) for controlling the flow of cooling water through a water/air exchanger above each table half. Commercial HEPA filters pull room air into the box. \textbf{(b)} The SrI vacuum chamber is contained in a black box to protect it from stray light and ensure thermal homogeneity. Heat sources that are temperature controlled are shown in red: Zeeman window (ZW), Zeeman slower (ZS), and MOT coils (MC). Temperature control of vacuum viewports is shown in blue: water temperature controlled copper tubes around the top and bottom 6" CF flanges (CF) and thermoelectric cooler (TEC) controlled 2.75" CF viewport faces (VF). \textbf{(c)} The temperature at the location of the atomic sample is directly measured using a thin-film platinum resistance thermometer (TFPRT) sensor and compared to a model derived from ray-tracing and temperature sensors mounted on the chamber, verifying the stability afforded by our extensive thermal control. The measurements shown are binned into 10 minute intervals. \textbf{(d)} At several hours averaging times the TFPRT shows sub-mK level stability. The difference between the TFPRT and model shows similar temperature stability, providing verification that long-term fluctuations in the temperature experienced by the atoms are accurately captured by the ray-tracing model.}
\label{figure_bbr_main}
\end{figure}

The frequency shift induced by blackbody radiation (BBR) is the largest systematic shift and a dominant source of uncertainty in state-of-the-art optical lattice clocks. Aside from cryogenic systems \cite{ushijima2015cryogenic}, the BBR-induced clock shift for strontium is approximately $5\times 10^{-15}$ at room temperature. The BBR shift of a thermal electric field distribution characterized by a temperature, $T$, may be expressed as:

\begin{equation}
\Delta \nu_{BBR}(T) =  \nu_{stat} \bigg( \frac{T}{T_0} \bigg ) ^4 +  \nu_{dyn} \bigg [ \bigg ( \frac{T}{T_0} \bigg ) ^6 + \mathcal O \bigg ( \frac{T}{T_0} \bigg ) ^8 \bigg ]
\end{equation}

\noindent
where $\nu_{stat} = -2.13023(6)$ Hz \cite{middelmann2012high}, $\nu_{dyn}=-148.7(7)$ mHz \cite{nicholson2015systematic}, and $T_0 = 300$ K. 

Characterizing the room temperature BBR shift at the low $10^{-18}$ level requires absolute knowledge of the thermal environment of the atoms to within $50$ mK. To date this challenging technical requirement has been met using two approaches. In the JILA SrII clock, NIST-calibrated in-vacuum sensors were used to directly measure the thermal environment of the atoms to an uncertainty of 5 mK ~\cite{nicholson2015systematic}. In the Yb OLC at NIST an in-vacuum radiation shield was characterized using precision thermometry and thermal modeling~\cite{beloy2014atomic}. Here we take components of both approaches by actively stabilizing and monitoring the thermal enclosure of the atoms while also utilizing in-vacuum thermometry.

Our primary objective is to create a frequency reference with low $10^{-18}$ level systematic uncertainty that does not require point-by-point corrections to attain a similar instability.  A homogeneous thermal environment is essential for accurate and precise characterization of the surroundings with a finite array of temperature sensors and for avoiding complications which arise when evaluating the AC BBR correction of a non-thermal spectrum of the electric field driven by temperature gradients \cite{nicholson2015systematic}.  Furthermore, actively maintaining a constant operating temperature leads to a more stable and reproducible BBR shift that is crucial for post-correction-free operation of the clock at the low $10^{-18}$ level. 

For our evaluation of the BBR shift in SrI, we utilized both a Monte-Carlo ray-tracing based thermal model and an in-vacuum thermal probe to determine the temperature at the atoms in two complementary ways. One method involves actively stabilizing and carefully measuring the thermal environment around the vacuum chamber using 16 servo loops and over 30 sensors.  The temperature at the atoms can then be computed from this array of temperature monitors using a detailed thermal model based on ray tracing as described in Appendix \ref{section_thermal_modeling}. We then verify the first approach by installing a PT100 thermal sensor in-vacuum to directly measure the BBR environment. Excellent agreement between the thermal model and the in-vacuum temperature measurement demonstrates that our model accurately predicts changes in the temperature experienced by the atoms from our external vacuum sensors readings at sub-mK levels. Our technical efforts focus on (1) active temperature stabilization of our system, (2) creating an array of accurate temperature sensors to characterize the thermal boundary conditions, and (3) calibration of an in-vacuum temperature probe. To this end, we improve the overall thermal homogeneity of JILA SrI by more than a factor of 10, limiting thermal gradients across the chamber to less than 100 mK.

We construct a model of the temperature $T_{i}$ at the location of the atoms, modeled by a small spherical surface $i$, by mapping the measured temperatures on the surrounding surfaces of the vacuum chamber to $T_i$ by:

\begin{equation}
T_{i}^4 = \sum_j F_{i \rightarrow j} T_j^4,
\label{equation_atom_temperature}
\end{equation} 

\noindent where the index $j$ enumerates the different surfaces of the vacuum chamber surrounding the atoms and $F_{i \rightarrow j}$ is the exchange factor defined as the fraction of total energy emitted by surface $i$ that is absorbed by surface $j$ directly or by reflection \cite{modest2013radiative}. A description of our exchange factors is given in Appendix \ref{section_thermal_modeling}. In the limit where the chamber temperature is nearly uniform, the uncertainty in $T_{i}$ can be expressed as

\begin{equation}
\delta T_{i} \approx \sum_j F_{i \rightarrow j} \delta T_j.
\label{equation_effective_uncertainty}
\end{equation}

\noindent
This limit is important for understanding how to prioritize temperature control of the experimental apparatus: surfaces with large exchange factors dominate the thermal environment of the atoms and are subsequently the most important to control. A useful heuristic for identifying the largest exchange factors, and therefore the most critical surfaces for thermal control, is to find the most highly emissive surfaces in the vacuum system subtending the largest solid angles at the location of the atoms. Often, and in our case, these surfaces are large vacuum windows.

To achieve the required temperature uniformity and stability, we began by stabilizing the air temperature around the experimental apparatus (Fig.~\ref{figure_bbr_main}a). The optical table is partitioned to isolate the vacuum chamber from the auxiliary laser systems and the entire table is surrounded by curtains to isolate it from the room air temperature fluctuations. Around the vacuum chamber a black box is installed to provide additional thermal homogeneity. Both sides of the table are actively temperature stabilized by controlling their inlet air temperature. Our chamber has several heat sources including the MOT coils, Zeeman slower, heated Zeeman window, and oven (Fig.~\ref{figure_bbr_main}b). Each heat source is accompanied with water cooling such that the relative heating/cooling rate allows us to actively control the steady-state temperature of each.

The equilibrium temperature at the atoms is determined from Eqns.~\ref{equation_atom_temperature} and \ref{equation_effective_uncertainty}, which are dominated by the contributions from the fused silica viewports due to their high emissivity and large solid angles with respect to the atomic sample.  Therefore, stabilization and accurate measurement of the viewport temperatures is of paramount importance. To stabilize their temperature, we fix the thermal boundary conditions of all viewports (Fig.~\ref{figure_bbr_main}b). The stainless steel flanges of the 6" viewports are wrapped in copper tubing carrying temperature controlled water to stabilize the temperature of the steel around the glass.  A copper ring with two embedded temperature sensors and lined with thermally conductive silicon matting is pressed against the outer edge of the glass (leaving a small aperture in the middle for optical beams to pass through) to achieve thermal homogeneity across the surface and allow direct monitoring of the glass temperature. All 2.75" viewports are controlled using custom, TEC temperature controlled aluminum attachments backed by water cooling.

This unique approach to active temperature control creates a stable thermal environment where temperature-controlled viewports and flanges act as thermal reservoirs with boundary conditions set by the servo setpoints. All setpoints for temperature control are set to 22$^{\circ}$C. To characterize the homogeneity of this thermal environment, 30 witness sensors with accuracy of 50 mK were placed around and on the chamber. Each viewport has an independent witness sensor, with the larger 6" viewports each having two. With the data from this sensor array and a detailed 3D model of the vacuum chamber as inputs to our thermal model, one can now compute the temperature seen by the atoms using Eqn. \ref{equation_atom_temperature}.

Upon completion of clock operation, an in-vacuum thermal probe is inserted into the vacuum chamber in order to directly measure the temperature at the position of the atoms. Placement of the probe at the position of the atoms is ensured by aligning the sensor to overlap with both the clock and the Zeeman slowing laser beams and the temperature is measured under identical run conditions to clock operation. This measurement allows for both the direct verification of the thermal model connecting temperatures measured on the vacuum chamber to the temperature seen by the atoms as well as the reduction of uncertainties associated with the emissivity of different vacuum components. The in-vacuum thermal probe consists of a 60 cm evacuated glass tube with a calibrated ($\pm 1.4$ mK) PT100 sensor epoxied at the end. The sensor design and calibration details can be found in Appendix \ref{section_calibration}. After installing the in-vacuum sensor, a small static offset of 19.1 mK was discovered between the temperature measured by the zero power resistance of the probe and the temperature derived from the thermal model using parameters given in literature, shown in Fig.~\ref{figure_bbr_main}c. This offset is attributed to a limited knowledge of material emissivities as well as calibration uncertainties of the thermistors (50 mK) of the majority of the sensor array. The in-vacuum probe measurement allows for the characterization of any static offset between the thermal model and the directly measured temperatures. Active stabilization of the thermal environment allows this offset to remain stable over clock operation where the thermal model successfully captures all fluctuations in the temperature to better than 1 mK over $>$10,000 s of averaging as indicated by the Allan deviation of the difference between the model and the measured temperature shown in Fig.~\ref{figure_bbr_main}d. All parameters of the thermal model and a more detailed discussion of the construction of the model can be found in Appendix \ref{section_thermal_modeling}.

The total temperature uncertainty quoted in Table \ref{tab:bbr} has contributions from the calibration uncertainty of the in-vacuum sensor, immersion error, self heating, modification of the thermal environment by the sensor (insertion error), and statistical error of the agreement between the probe and thermal model.
Immersion error in the SrI chamber was measured by changing the base flange temperature and monitoring the in-vacuum sensor, giving a slope of (0.65 $\pm$ 0.62) mK/K.
During clock operation, this flange is controlled to 0.2 K, and so we assign the total immersion error uncertainty to be the quadrature sum of the overall offset and the coefficient uncertainty, arriving at a 1.8 mK uncertainty. 
Insertion error is bounded to 1.5 mK from previous work comparing the temperature measured at one thermistor to the temperature measured at another as the position of the second is translated away from the chamber center \cite{nicholson2015systematic}. The total uncertainty in the temperature of blackbody radiation seen by the atoms is then evaluated to be $\delta T = 2.9$ mK corresponding to an uncertainty of 2.0$\times 10^{-19}$. Uncertainty in the combined static and dynamic BBR coefficients accounts for the atomic response contribution to our BBR uncertainty, at $1.49\times10^{-18}$.

\begin{table}[h]
	\centering
	\caption{Atomic Temperature Uncertainty}
	\label{tab:bbr}
	\begin{tabular}{  l  c r }
		\hline
		Shift & Correction (mK) & Uncertainty (mK) \\ \hline
		Sensor calibration & $0$ & 1.4 \\ 
		Self heating & -1.4 & 0.3 \\ 
		Immersion error & $0$ & 1.8 \\ 
		Sensor - model & $20.5$ & 1.0 \\
		Insertion error & $0$ & 1.5 \\ 
		\textbf{Total} & \textbf{19.1} & \textbf{2.9} \\
	\end{tabular}
\end{table}

\subsection{Density Shift}

The high-degree of stability demonstrated by SrI and more generally by optical lattice clocks is due to the ability to interrogate thousands of atoms simultaneously and read-out the measured clock frequency with high signal-to-noise at the limit set by quantum projection noise (QPN)~\cite{Oelker}. However, the presence of multiple spin-polarized atoms per lattice site introduces systematic frequency shifts due to $p$-wave interactions. The different triplet collision channels between ground and excited atoms have been shown to have different scattering lengths which subsequently produce density-dependent differential clock shifts. These effects have been studied and characterized in $^{87}$Sr \cite{martin2013quantum,zhang2014spectroscopic}.

During clock operation, the gas is spin-polarized by optical pumping into either the $\vert \frac{9}{2}, \frac{9}{2}\rangle $ or the $\vert \frac{9}{2}, -\frac{9}{2}\rangle $ ground hyperfine state before interrogation in order to suppress frequency shifts due to $s$-wave collisions. However, $s$-wave interactions can still occur if impurity atoms remain present after the optical pumping process. To mitigate this effect - and suppress temporal variation of this effect due to fluctuations in the efficiency of optical pumping - we perform clock spectroscopy on impurity nuclear spin states and optimize the optical pumping process to reduce the population in other nuclear spin states below the detection threshold of our system. In this regime, the dominant density-dependent frequency shift is a collective $p$-wave interaction between identical fermions.

To reduce the QPN-limited clock stability to the low-$10^{-17}/\sqrt{\tau}$ level, we initially prepare more than $1000$ atoms in the optical lattice. For these atom numbers in the deepest optical lattice depths $\sim$180 $\text{E}_\text{r}$, on-site densities are high enough that sub-percent-level changes in the trap depth or atom distribution can produce variations of the shift at the $10^{-18}$ level. The effect of these fluctuations is suppressed to the $10^{-19}$ level and below by lowering the lattice depth to 45 $\text{E}_\text{r}$ thereby reducing the peak atomic density during clock interrogation. Vertical orientation of the 1D lattice ensures suppression of tunneling due to the difference in gravitational potential energy between lattice sites.

In addition, fluctuations in the atomic density and the resulting density shift can be driven by changes in the laser-cooling process which transfers the atoms from the narrow-line red MOT into the lattice. This loading process is influenced by the relative frequency, intensity, polarization, and alignment stability of the red MOT lasers. To control these processes, the lasers are stabilized to an ultra-stable cavity to a linewidth of 1 Hz. Temperature stabilization of the experimental apparatus - as discussed in the previous section - serves the dual purpose of stabilizing the alignment of the red MOT and enables robust, alignment-free operation of the clock over many consecutive months. After implementing these measures, we find both the sample temperature and the spatial distribution of atoms in the lattice are insensitive to daily linewidth-level frequency drifts and alignment drifts. Linear scaling of the temperature with the loading lattice depth indicates that final temperatures in the lattice are dominated by the large AC Stark shift of the 813 nm lattice on the $^3P_1$ cooling transition. Since the lattice depth is actively stabilized, the atomic distribution within the lattice remains static over time leading to a density shift with stability at the $10^{-19}$ level.

\begin{figure}[h]
\centering
    \includegraphics[width=0.50\textwidth]{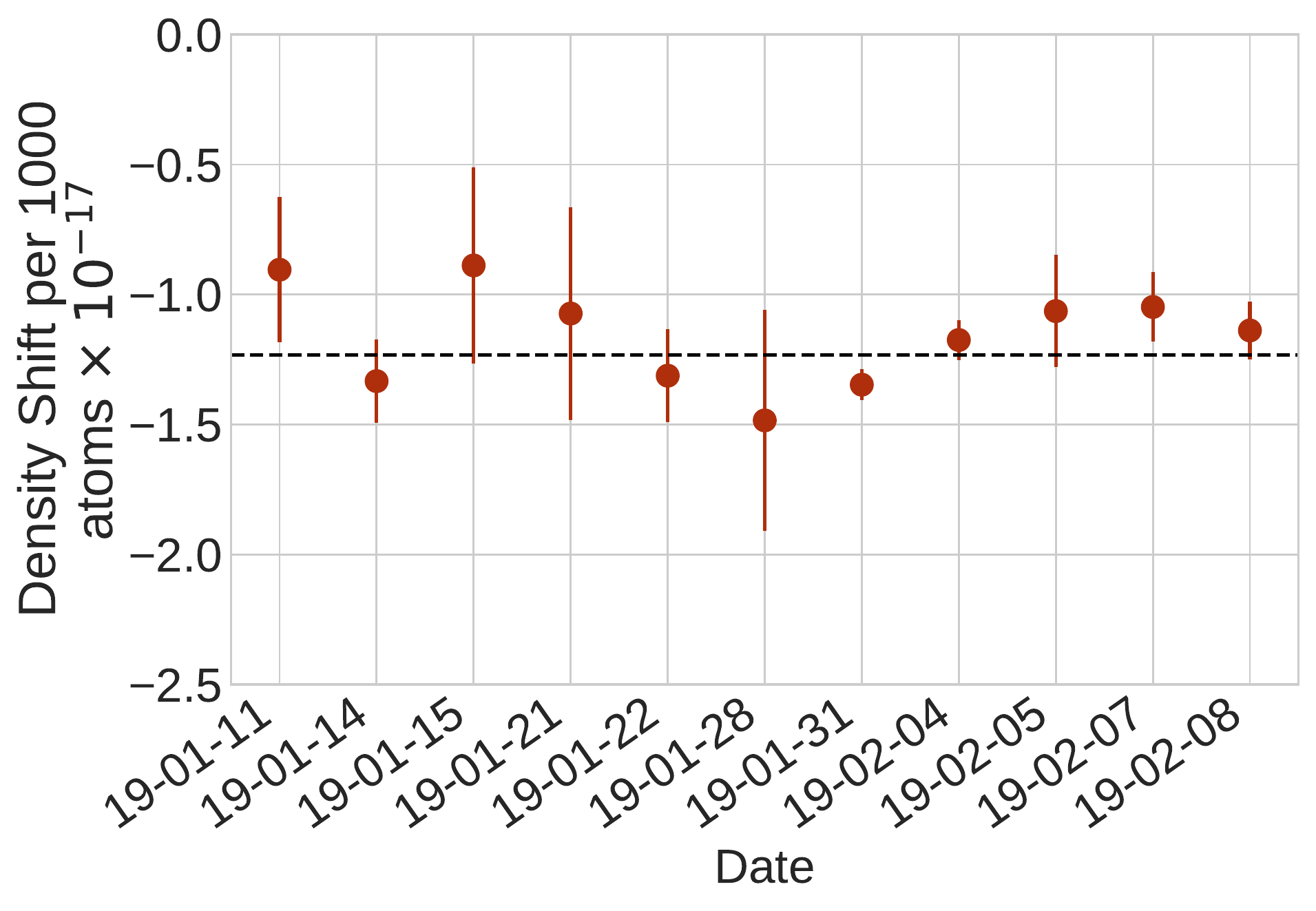}
    \caption{\textbf{Density shift evaluations.} Density shift measurements were performed over four weeks at the same trap conditions. The weighted mean of all measurements is shown with a dashed, black line, with a reduced chi-squared of 1.07.}
\label{fig:Dens_DC_plots}
\end{figure}

To measure and confirm this stability, the density-dependent shift is determined by modulating the atom number by varying the loading time of the blue MOT which then varies the atomic density in the lattice. Care is taken to ensure that, at the highest atom numbers, the spectroscopic lineshape remains Fourier-limited with high peak excitation fraction unblocked by a many-body excitation gap \cite{martin2013quantum}. 
A model of the density shift which is linearly dependent on atom number is applied to extrapolate the measured shift to the atom number in the lattice on any particular experimental realization. 
To demonstrate the stability of the shift, repeated density shift evaluations were performed at our nominal clock lattice depth of $45 \: \text{E}_{\text{r}}$ over the course of several weeks, with a weighted standard error of $3.9\times 10^{-19}$ and $\chi^2 = 1.07$ for all these measurements (Fig.~\ref{fig:Dens_DC_plots}). 

Finally, by verifying that the radial and axial trap frequencies in the lattice scale with trap depth as expected for a thermal gas, we confirm the shift scaling of $\Delta\nu_{density}\propto U^{5/4}$ as reported in \cite{MattSwallows}. Fig.~\ref{fig:DensityACstark}a shows the result of this evaluation where each point has error bars evaluated to the low $10^{-18}$ level. The data fits well to a model with the functional form: $a + b U^{5/4}$, where $U$ is the trap depth and $a$ and $b$ are fit parameters. Though we independently measure the density shift for a particular lattice depth each day during clock operation, this scaling becomes highly useful when evaluating shifts at several trap depths, as done during the AC Stark evaluation described in the next section. 

\subsection{Lattice AC Stark}

To eliminate Doppler shifts, the atoms are tightly confined in a one-dimensional optical lattice to perform clock spectroscopy in the Lamb-Dicke limit. This confinement induces a differential AC Stark shift between the ground and excited clock states. To minimize this effect, we operate our lattice near the so-called magic frequency where the differential polarizability between the ground and excited states is nearly zero.

\begin{figure}[!htb]
\centering
\includegraphics[width=0.9\textwidth]{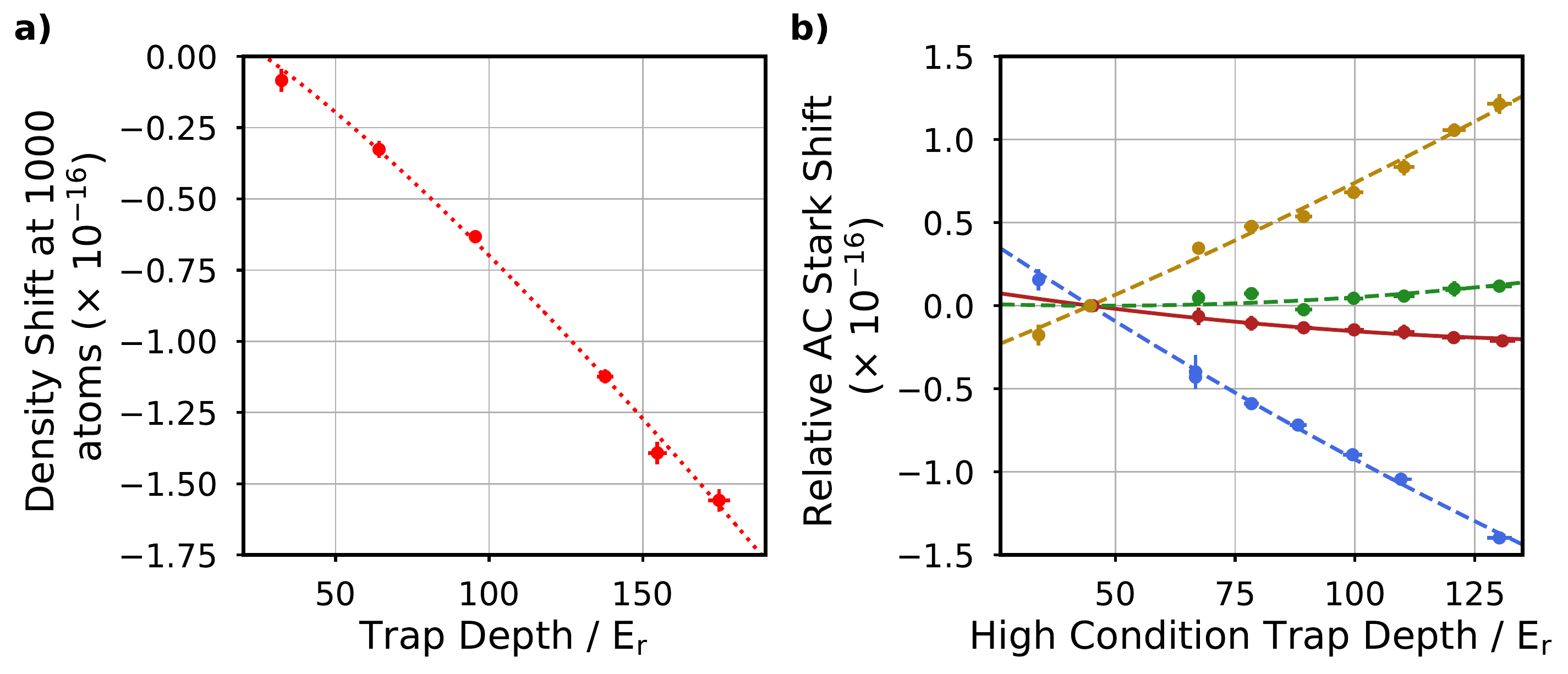}
\caption{\textbf{Scaling of the light shift with trap depth $U$ (a)} Density shift evaluated at different trap depths, scaled to a shift with 1000 atoms. The dashed red line is a fit to the data with the expected scaling of the shift as $U^{5/4}$. \textbf{(b)} Lattice Stark shifts measured relative to a trap depth of 45$E_r$. Four different lattice frequencies are shown: $\nu_L =$ 368.55452610 (blue), 368.55447610 (red), 368.55446610 (green), and 368.55442610 THz (gold). The three curves with dashed lines are independently fit using the model in Eqn.~\ref{eq:thermalmodel}, obtaining a weighted mean of their $\beta^*$'s. Using this $\beta^*$, the red curve is then fit for $\alpha^{*}_{\text{clock}}$, fully characterizing our AC Stark shifts for clock operation at the red curve. Vertical error bars are obtained from fits to the Allan deviation of each evaluation, extrapolated to the total measurement time, and scaled by the lever arm of the measurement. Horizontal error bars are uncertainties on our determination of the trap depth, obtained from axial sideband scans.}
\label{fig:DensityACstark}
\end{figure}

To constrain the lattice Stark shift at the $10^{-18}$ level, recent work has highlighted the importance of accounting for higher-order effects such as hyperpolarizability and the magnetic dipole and electric quadrupole terms ~\cite{KatoriACStark,MariannaPol}. These manifest in light shifts scaling with nonlinear powers in trap depth. Recent work \cite{BrownACstark} has demonstrated that, given a thermal scaling of axial and radial modes with trap depth $U$, these nonlinear shifts can be greatly simplified to a linear and quadratic term, expressed as:

\begin{align} \label{eq:thermalmodel}
    \frac{\nu_{LS}}{\nu} \simeq \alpha^* U + \beta^* U^2.
\end{align}
Eqn.~\ref{eq:thermalmodel} shows that characterization of the lattice light shift requires the measurement of experiment-specific coefficients $\alpha^*$ and $\beta^*$. A discussion of the uncertainty associated with the use of this model in the SrI system can be found in Appendix \ref{section_ac_stark}. Note that $\alpha^{*}$ and $\beta^{*}$ not only depend on atomic coefficients, but also on the atomic distribution in both the radial and axial directions. As such, care should be taken to evaluate these coefficients under reproducible sample temperatures and lattice trapping frequencies. 

To evaluate the lattice light shift a series of differential measurements of the light-induced frequency shift between a series of lattice depths ranging from 34 E$_{\text{r}}$ to 141 E$_{\text{r}}$ and a reference depth of 45 E$_{\text{r}}$ is performed. These measurements are repeated for four different lattice frequencies, $\nu_L$, encompassing both positive and negative detunings from the magic frequency. This enables accurate determination of $\nu_L$ dependent $\alpha^*$'s and a $\nu_L$ independent $\beta^*$. The resulting data is shown in Fig. \ref{fig:DensityACstark}b. We perform this evaluation by relying on measurement precision as opposed to a large measurement lever arm, enabling us to stay close to the clock operational trap depths and atomic distributions. To remove trap depth dependent density shifts from the lattice light shift evaluations, we perform a series of density shift measurements over a range of lattice depths to which we fit a $U^{5/4}$ model, shown in Fig. \ref{fig:DensityACstark}a. The residual uncertainty of the $U^{5/4}$ model fit to the density shift data in Fig. \ref{fig:DensityACstark}a is then propagated to the lattice light shift measurements and added in quadrature with the statistical uncertainty of the measurement.

Lattice lights shifts are evaluated at four different lattice frequencies, shown by the four different curves in Fig. \ref{fig:DensityACstark}b. The determination of the lattice light shift at the operational lattice frequency proceeds in two steps. First, data from three different lattice frequencies (blue, green, and yellow points) is used to determine the wavelength-independent quantity $\beta^{*}$ by least-squares fitting to the model in Eqn. \ref{eq:thermalmodel}. Taking the weighted mean and weighted standard error of the mean of the three $\beta^*$ values from these fits results in $\beta^* = 1.93 (20) \times10^{-21}$. Second, $\beta^*$ is then used to fit $\alpha^{*}_{\text{clock}}$ to the data taken at the operational lattice frequency (red points), resulting in $\alpha^{*}_{\text{clock}} = -5.61(22)\times 10^{-19}$. The total AC Stark uncertainty is found in Tab. \ref{tab:Stark_budget} and includes uncertainties coming from $\alpha^{*}$ and $\beta^{*}$ as well as additional contributions from a 2\% uncertainty in trap depth and a model uncertainty. In total, this procedure allows the evaluation of the lattice light shift with a total uncertainty of $1.2 \times 10^{-18}$.

To maintain this low uncertainty, we eliminate temporal variations of the lattice light shift. To accomplish this, the trapping light is spectrally filtered and frequency stabilized. First, broadband amplified spontaneous emission (ASE) from the high power Ti:sapphire laser which generates the trapping light is suppressed by reflecting the laser light off two volume Bragg gratings each providing in excess of 30 dB of suppression of ASE power outside the $\sim$10 GHz wide passband. The Ti:sapphire laser is then phase-locked to an Er:fiber frequency comb which is stabilized to a cryogenic silicon reference cavity~\cite{mateiSiCavity}. As a result, the drift of the lattice frequency is limited to the low and well characterized drift rate of the cavity (-7.4 Hz/day at $813$ nm), which produces negligible drift of the lattice Stark shift. With this stabilization scheme, measurement of the cavity frequency by routine clock operation allows determination of the absolute lattice frequency at the sub-Hz level. In addition, alignment of the lattice polarization to the bias magnetic field and alternately probing the stretched states suppresses the vector contribution and sensitivity to fluctuations in the tensor term. Fluctuations in the relative direction of the magnetic field and the polarization axis are additionally suppressed by the active temperature control of the experimental apparatus with residual background field fluctuations observed at the 1 mG ($100$ nT) level, corresponding to a shift constrained below the $10^{-19}$ level. To suppress drifts in the lattice intensity, the laser power is actively stabilized by monitoring the reflection of the incident light from the top surface of the vacuum window. To ensure consistent overlap of the in-going and retroreflected lattice beams, a small amount of power from the retroreflection which is transmitted back through the optical fiber used for beam delivery is monitored to ensure stability of the lattice alignment. Finally, examining Eqn. \ref{eq:thermalmodel}, we see that the light shift is sensitive to fluctuations in the atomic distribution. As covered in the discussion of the density shift, several steps were taken to ensure robustness of both temperature and density and this is reflected in the high reproducibility of the density shift in Fig.~\ref{fig:Dens_DC_plots}.

\begin{table}[H]
	\centering
	\caption{Lattice light shift uncertainty contributions}
	\label{tab:Stark_budget}
	\begin{tabular}{ l l r}
		\hline
		Parameter & Value (Uncertainty) & Uncertainty ($10^{-19}$)  \\ \hline
		$\alpha^*$ & -5.61(22) $\times 10^{-19}$ & 9.7 \\ 
		$\beta^*$ & 1.93(20)$\times 10^{-21}$ & 4.1 \\ 
		$U (E_r)$ & 45.0(9) &  3.5\\ 
		Model &  & 3.3\\
		\textbf{Total} &  & \textbf{11.6}\\ \hline 
	\end{tabular}
\end{table}

\subsection{DC Stark Shift}

Stray DC electric fields or patch charges on vacuum viewports can induce a frequency shift due to the differential DC polarizability ($\alpha_0$) between the two clock states. To directly measure this effect with the atoms, a pair of ring electrodes are placed on the top and bottom viewports which have the closest proximity to the atoms. Each ring consists of four copper quadrants which can each be independently biased. By applying a large lever arm of $\pm100$V to the correct combination of electrodes a shift can be measured along any of the three Cartesian axes.

Due to the quadratic dependence of the shift on the total electric field, the magnitude of the background field along each direction can be determined by performing a two-point measurement. We measure $\Delta\nu_+$, the frequency shift when a large field is applied along one direction and compared to the reference case where both electrodes are grounded. The field direction is then reversed and the frequency, $\Delta\nu_-$, is recorded. The direction of the background field, along with its corresponding field amplitude, can then be computed from these two measurements by noting that $\Delta\nu_{\pm}=-\frac{1}{2}\alpha_0(E_{a}\pm E_{r})^2$ for applied and residual fields $E_{a}$ and $E_{r}$. We observe a modest background field at the low $10^{-18}$ level in the vertical direction and no clearly resolvable field at the $10^{-20}$ level along the horizontal axes. By applying $(-4.2$V, $ +4.2$V) to the (top , bottom) electrodes, respectively, we cancel this background shift to an uncertainty of $2.5 \times 10^{-19}$.

\subsection{Second Order Zeeman Shift}

As described in Section 2, the Zeeman shift is cancelled to first order by averaging frequency measurements of the $m_F =\pm \frac{9}{2}$ clock transitions.  However, the Zeeman Hamiltonian contains a term with a quadratic dependence on the magnetic field which is not suppressed by this technique.  The 570 mG bias field applied to resolve the hyperfine structure of the clock states induces a second order Zeeman shift of $\approx 77$ mHz.  This shift is typically expressed as 

\begin{equation}
\Delta\nu_{B,2} = \xi\left(\Delta\nu_{B,1}\right)^2
\end{equation}

\noindent where $\xi$ is the quadratic Zeeman coefficient for $m_F =\pm \frac{9}{2}$ and $\Delta\nu_{B,1}$ is the first-order Zeeman splitting in Hz between the $m_F =\pm \frac{9}{2}$ transitions due to the applied bias field.  Previous uncertainty in the value of $\xi$ introduced a sizeable term in our error budget, motivating a more precise evaluation of the coefficient. 

The determination of $\xi$ is complicated by the fact that the observed splitting $\Delta$ between the $m_F =\pm \frac{9}{2}$ transitions also contains a contribution from the vector AC Stark shift $\delta\nu \approx 0.4$ Hz.  In analogy with  $\Delta\nu_{B,1}$, the vector AC Stark term is often thought of as arising from a synthetic magnetic field.  We evaluate $\xi$ in the limit where the applied bias field is much larger than this synthetic field so that changing the bias field magnitude does not significantly rotate the total effective field vector.  We validate that we are in the appropriate limit by measuring the lattice vector shift directly in an AC Stark evaluation to determine the magnitude of the synthetic field.  In this limit, the second-order Zeeman shift can be expressed as $\Delta\nu_{B,2} = \xi(\Delta -\delta v)^2$.  By performing differential measurements of the clock transition frequency $f_0$ at three different bias field values, we may determine both $\xi$ and  $\delta\nu$ from the following system of equations:
\begin{align}
\begin{split}
     f_{0_2} - f_{0_1} &= \xi(\Delta_2^2 -\Delta_1^2) - 2 \xi \delta v(\Delta_2 - \Delta_1) \\
     f_{0_3} - f_{0_2} &= \xi(\Delta_3^2 -\Delta_2^2) - 2 \xi \delta v(\Delta_3 - \Delta_2)
\end{split}
\end{align}

\begin{figure}[!htb]
\centering
\includegraphics[scale=0.5]{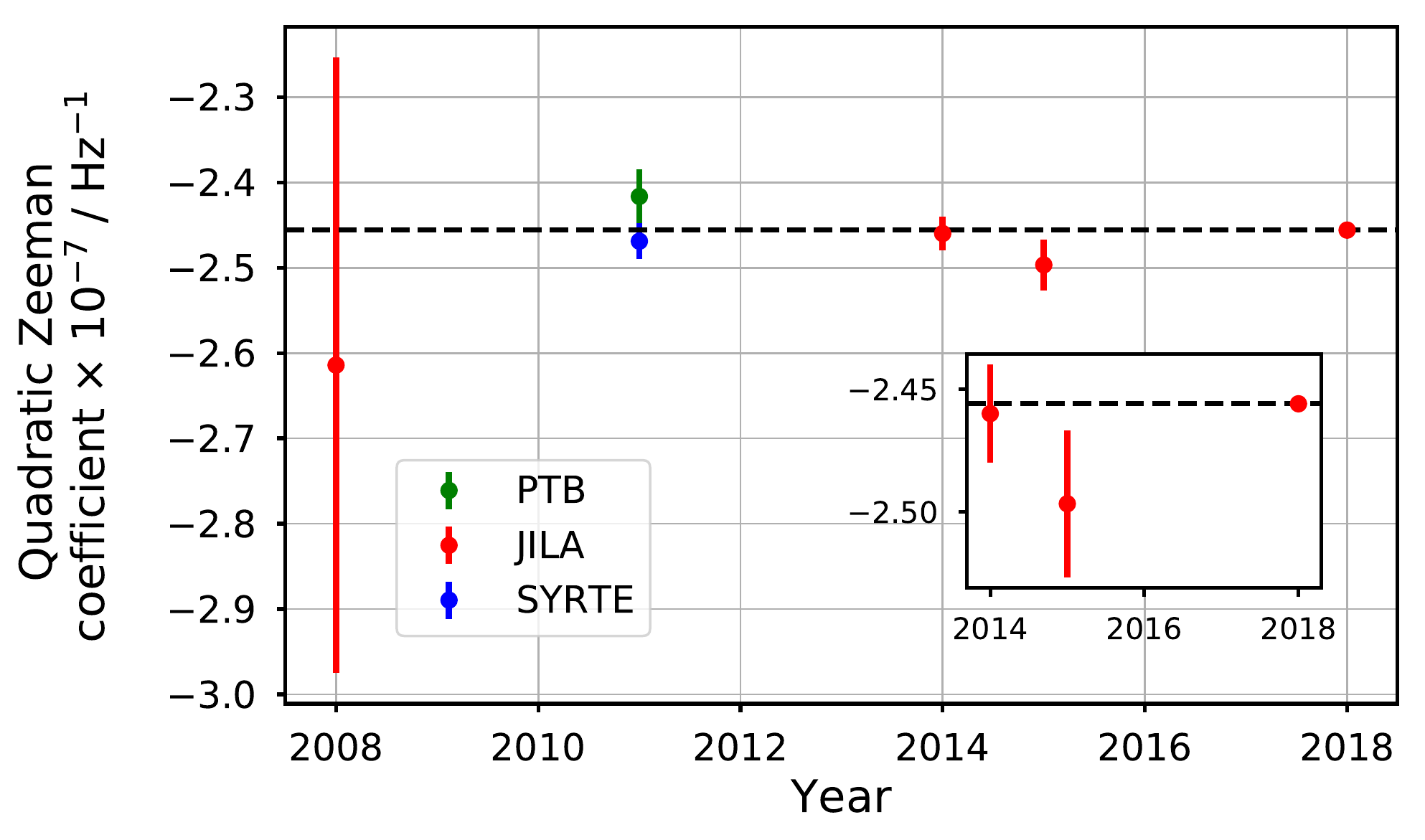}
\caption{\textbf{Evaluation of second order Zeeman coefficient.} A history of evaluations of the Sr second order Zeeman coefficient completed by the PTB \cite{Falke}, SYRTE \cite{SYRTEhyperpol}, and JILA \cite{GretchenCampbell,bloom2014,nicholson2015systematic} Sr OLCs. The dashed black line is a weighted mean of all six measurements and the inset shows the three most recent evaluations.}
\label{fig:Zeeman}
\end{figure}

This measurement is repeated at three different sets of bias field values all yielding consistent measurements of $\xi$.  We report the weighted mean of these three evaluations yielding a value of $\xi=-2.456(3)\times 10^{-7}$ Hz$^{-1}$. Fig.~\ref{fig:Zeeman} shows this result along with historical evaluations showing good agreement with previous measurements. Improved knowledge of this coefficient reduced the uncertainty associated with the second-order Zeeman shift by a factor of 5 to $2\times 10^{-19}$.

\subsection{Background Gas Collisions}

Two-body collisions between a cold $^{87}$Sr atom and a room temperature hydrogen molecule or a hot Sr atom emitted by the atomic beam source have the potential to cause a systematic shift. In an uncorrelated thermal gas, a high-energy collision which removes an initially trapped atom from the sample does not produce a systematic shift. However, atom-light coherence can be affected by collisions with low momentum transfer that produce phase-shifts while leaving the atom trapped. In the process of a collision, the ground and excited states shift differentially as a function of the distance between the Sr atom and its collision partner. When an atom is placed in a superposition of ground and excited states during clock interrogation, the integrated energy difference over the time of the collision produces an undesirable phase shift which appears as a systematic frequency shift varying linearly as a function of the background gas density.

Under ultrahigh vacuum conditions, this frequency shift can be well-approximated by considering collisions between $^{87}$Sr and hydrogen molecules \cite{MitroyZhang}. The SrI vacuum chamber is stainless steel and contains no getter pump, so the dominant background gas contribution is primarily molecular hydrogen gas. The SYRTE collaboration recently measured a background gas collisional shift of $(-3.0\pm 0.3)\times10^{-17}/\tau$ for a hydrogen-limited vacuum lifetime $\tau$~\cite{Jerome}. For the SrI clock, the lifetime of a dilute sample of $^{3}\text{P}_{2}$ atoms trapped in the quadrupole field of the MOT is measured to be 8.1(2) s after correcting for BBR-induced decay \cite{yasuda2004lifetime}. The measured lifetime of ground state atoms trapped in the optical lattice is the same as the lifetime in the quadrupole trap \cite{martin2013quantum} and leads to a background gas collisional shift of $(-3.7\pm 0.4)\times10^{-18}$.

In addition, two-body collisions with hot Sr atoms emitted from the atomic beam source may also cause a frequency shift. To evaluate this possible systematic shift, an atomic beam shutter is in place to block line of sight to the oven during clock interrogation. The oven flux is increased above the normal operating condition by a factor of 21, and interleaved interrogation of the clock frequency with and without the atomic beam shutter closed is performed. With this lever arm, no significant shift is observed at a measurement precision of $1.0\times10^{-18}$, therefore bounding the shift at normal operating conditions to below the $10^{-19}$ level. The shutter is not used for normal clock operation due to intermittent failure of the shutter in the closed position. No difference in atom number or lifetime in the trap is observed with the beam shutter closed versus open.

\subsection{Doppler Shift and AOM Phase Chirp}

For narrow line spectroscopy, Doppler shifts between the clock laser and atoms trapped in the optical lattice are eliminated by active optical path length stabilization~\cite{FNC}. Ideally, the surface which serves as a reference for the laser phase for clock interrogation is the same as the surface which serves as the phase reference for the optical lattice. However, when the clock laser intensity is suddenly changed from zero to a finite value to drive the clock transition, the response of the active stabilization loop experiences transient phase shifts which may impart systematic frequency shifts.

Simultaneous laser phase stabilization and atomic spectroscopy is accomplished with two AOMs -- one before an optical fiber that is the actuator for path length stabilization, and a second after the fiber that is the actuator for steering the laser frequency to the atomic resonance. In the 2013 systematic evaluation of the SrI clock the stabilization of the laser phase was accomplished by retroreflecting the zeroth order of the second AOM from the mirror to which the optical lattice is referenced while the first negative order was used for atomic spectroscopy. When the clock pulse turns on, this stabilization method produces a differential path length change between the zeroth- and first-diffracted orders due to thermal changes in the AOM crystal. In addition, both differential path length changes driven by air currents between the two diffracted orders and reflections from the tip of the optical fiber add phase noise above the level of the clock laser in the path length stabilization servo.

To circumvent the problems associated with thermal effects in the AOM crystal, fiber tip reflections, and differential path length noise, the path length stabilization scheme is revised such that the first negative diffracted order from the AOM after the fiber is now used for both spectroscopy and path length stabilization -- eliminating all differential optical path between the path length stabilization light and the light used for atomic spectroscopy. Mounting a wedged beam sampler to the back surface of the optical lattice retroreflector mirror allows $10\%$ of the incident clock light to be used for path length stabilization. Additionally, Rabi pulses are implemented by first turning on clock light at a large (1 MHz) detuning for 7 ms and then performing a 3.4 ms ramp onto resonance. This method allows the clock intensity servo to settle before spectroscopy and eliminates all thermally-induced differential path length transients in the AOM crystal. Furthermore, phase transients in the path length stabilization in-loop error signal that arise when operating the servo in a pulsed fashion now occur when the clock light is detuned and the atomic sensitivity function is zero.  Eliminating sensitivity to these transients removes a potential source of systematic frequency offsets~\cite{falkeFNC}. The fractional frequency shift due to the linear frequency ramp to resonance is estimated with the Landau-Zener transition probability for the parameters above and is calculated to be $2\times10^{-22}$ for a 600 ms pi-pulse.

\subsection{Line Pulling}

In addition to collisional shifts, off-resonant excitation of an impurity spin population can produce a systematic line pulling frequency shift. The impurity spin population is determined by performing clock spectroscopy on impurity spin states and detecting the number of atoms promoted to the excited state. The total impurity spin population is determined to be better than 99.5\%, limited by the detection threshold of the fluorescence measurement. The line pulling effect is subsequently bounded by taking the maximum population in the $\pm|5/2\rangle$ and $\pm|7/2\rangle$ states after dark-state optical pumping and calculating their maximum contributions to the excitation fraction observed on the $\pm|9/2\rangle$ transition. For a 600 ms pi-pulse and a 62 Hz splitting between the $|9/2\rangle$ and $|7/2\rangle$ clock transitions, the maximum off-resonant excitation is given by the ratio of Clebsch-Gordon coefficients squared and the off-resonant Rabi frequency: $(\frac{0.49}{0.82})\times (\pi/0.6)^2 / ((2\pi\times 62)^2 + (\pi/0.6)^2) \approx 1\times 10^{-4}$. Combining the upper bound of the atom number in impurity spin states ($0.5\%$) with this excitation fraction and converting to frequency units, an upper bound on the line pulling effect is set at the $6\times 10^{-22}$ level.

\subsection{Servo Error}

The servo error systematic arises from the linear frequency drift of the clock laser that serves as a local oscillator, or from a drifting background magnetic field, producing a systematic bias of the excitation fraction error away from the desired lock point. To minimize this effect, a PI$^2$D digital servo is used for locking the clock laser to the atomic transition. The servo loop is tuned by optimizing the attack time of the lock error signal with respect to the steering control signal. As a result, for a measurement time of 40,000 seconds, a mean servo error of $-6\times10^{-19}$ is recorded, which averages to an uncertainty of $9\times10^{-20}$.

\begin{table}[]
\centering
\caption{SrI uncertainty table}
\begin{tabular}{ l c c } \hline
\label{tab:uncertainties}
Systematic        & Shift ($10^{-18}$) & Uncertainty ($10^{-18}$) \\ \hline
BBR (environment) & -4974.1      & 0.2                     \\
BBR (atomic)      & 0        & 1.5                     \\
Density           & -12.3      & 0.4                       \\
Lattice AC Stark  & -21.3         & 1.2                      \\
DC Stark          & 0         & 0.3           \\
Probe AC Stark    & 0           & \textless{}0.1           \\
1st order Zeeman  & 0        & \textless{}0.1           \\
2nd order Zeeman  & -176.9       & 0.2                      \\
2nd order Doppler & 0         & \textless{}0.1           \\
Servo error       & 0          & 0.2           \\
Line pulling      & 0        & \textless{}0.1           \\
Background gas    & -3.7      & 0.4                      \\
AOM phase chirp   & 0         & \textless{}0.1           \\ \hline
\textbf{Total}    & \textbf{-5188.3}     & \textbf{2.0}           
\end{tabular}
\end{table}

\section{Summary}

In summary, we demonstrate a significant advance in the characterization of strontium optical lattice clocks resulting in a low systematic uncertainty of $2.0\times10^{-18}$. In conjunction with its high uptimes as well as the highly predictable frequency evolution of the cryogenic silicon cavity, such a clock will be a core component of an optical timescale. Indeed, the JILA SrI clock has already been used in an atom-cavity frequency intercomparison which highlights the long term stability of cryogenic crystalline cavities for timescales applications~\cite{Milner} and the search for time-variation of fundamental constants~\cite{Kennedy}. As illustrated in Fig.~\ref{fig:AtomCorr}, fluctuations in systematic offsets are bounded below $4 \times 10^{-19}$ over $10^4$ seconds of operation. We additionally detail a powerful new technique to combat blackbody radiation shifts that provides a stable thermal environment in which in-vacuum thermometry is only needed as calibration of sophisticated temperature modeling. As clocks are pushed into the next decade of accuracy, the ability to remove the temporal variation of systematics will be a highly powerful tool. With the main limitations to our uncertainty being insufficient knowledge of atomic coefficients, this system sets the path for developing strontium optical lattice clocks at the $10^{-19}$ level.

\section{Acknowledgements}

This work is supported by the National Institute of Standards and Technology (NIST), the Defense Advanced Research Projects Agency (DARPA), the Air Force Office of Scientific Research Multidisciplinary University Research Initiative, and the National Science Foundation (NSF) JILA Physics Frontier Center (NSF PHY-1734006). E. Oelker and C. J. Kennedy are supported by a postdoctoral fellowship from the National Research Council. We acknowledge technical contributions from J. Scott, S. Kolkowitz, and J. Uhrich and useful scientific discussions with A. Goban, R. B. Hutson, and C. Sanner. We thank A. Ludlow, his group, and C. Sanner for careful reading of this manuscript. We thank J\'er\^ome Lodewyck and SYRTE for providing the background gas collisional shift coefficient.

\appendix
\section*{Appendix}

\section{Thermal Environment Evaluation}

\subsection{Thermal Modeling}
\label{section_thermal_modeling}

Radiative heat transfer between specular greybody radiators is a well-studied problem in a variety of systems ranging from ovens to satellites \cite{modest2013radiative}. To characterize the thermal radiation experienced by the atoms we must understand the radiative contributions from each surface of the vacuum chamber. We start by modeling the atoms as a small spherical surface, hereafter referred to as the probe, in thermal equilibrium with the vacuum chamber. The temperature $T_i$ of this surface, or equivalently of the radiation bathing the atoms, is given by

\begin{equation}
    \epsilon_i \sigma T_i^4 A_i = \sum_j \epsilon_j  \sigma A_j F_{j \rightarrow i} T_j^4
    \label{equation_exchange_factor_unsimplified}
\end{equation}

\noindent
where $j$ enumerates the surrounding vacuum chamber surfaces, $\epsilon_j$ is the surface's emissivity, $\sigma$ is the Stefan-Boltzman constant, $A_j$ the surface area, and $F_{j \rightarrow i}$ the exchange factor. The exchange factor $F_{j \rightarrow i}$ is defined as the fraction of total energy emitted by surface $j$ that is absorbed by surface $i$, either directly or after reflection from any intermediate surfaces. In this language, Equation \ref{equation_exchange_factor_unsimplified} is telling us that at thermal equilibrium, the total power radiated by our modeled atom surface must be equal to the total incoming power from the surrounding vacuum chamber. Similarly, the exchange factors in Equation \ref{equation_exchange_factor_unsimplified} tell us what portion of radiated power from each surrounding surface $j$ is incident on the probe surface.

Using our understanding of the system's behavior at thermal equilibrium we can simplify Eqn.~\ref{equation_exchange_factor_unsimplified} significantly. First, energy conservation requires that $ \sum _j F_{i \rightarrow j} = 1$, meaning all emitted energy must be absorbed by the surrounding surfaces. Second, reciprocity demands that $\epsilon_i A_i F_{i \rightarrow j} = \epsilon_j A_j F_{j \rightarrow i}$ \cite{modest2013radiative}. This can be understood by considering all of the paths that emitted BBR (rays) can propagate between surfaces $i$ and $j$ - whether the ray goes from $i$ to $j$ or $j$ to $i$ the paths connecting the surfaces are the same. With these two conditions we can simplify Eqn.~\ref{equation_exchange_factor_unsimplified} as

\begin{equation}
T_{i}^4 = \sum_j F_{i \rightarrow j} T_j^4.
\label{equation_effective_temperature}
\end{equation} 

\noindent
Note that the probe temperature no longer depends on the probe surface's emissivity - consistent with Kirchoff's radiation law. Also note that the exchange factors now consider radiation propagating from the probe surface. Once we have the exchange factors we then have the connection between the array of sensors on our vacuum chamber and the temperature experienced by the atoms.

Calculation of the exchange factors is performed using Monte Carlo ray tracing. This begins with a 3D CAD replica of our vacuum system which is broken into several pieces to address the different temperatures, emissivities, and reflective qualities (specular and diffusive) of each chamber component. From the location of the atoms we propagate rays with random orientations to evenly sample the 4$\pi$ steradians around the atoms. For each ray we find the intersection of that ray with the vacuum chamber boundary. Each surface has a probability to absorb (emissivity), diffusively reflect, or specularly reflect the incident ray. If a ray is absorbed, the location and surface is recorded. For reflected rays, the new ray direction is found and the process repeated until every ray has been absorbed by a boundary surface. We perform this process for 50 million initial rays to ensure sufficient ray intersections with all surfaces. Specifically, this ensures that the least intercepted surface, the oven nozzle, absorbs more than $1000$ rays for sufficient simulation convergence. From this record we calculate what fraction of the total number of rays are absorbed by each surface. This fraction yields the exchange factor for a given component of the vacuum chamber.

The emissivity and specular reflectivity values used in the ray tracing analysis are given in Table~\ref{tab:emissivities} and Table~\ref{tab:reflectivities} respectively. Uncertainties of these parameters may result in the small static offset between the temperature determined by a ray-tracing model and the in-vacuum sensor; however, using the calibrated sensor to account for this offset removes the need to propagate uncertainties associated with both emissivities and reflectivities. Our calculated exchange factors are given in Table \ref{tab:exchange_factors}. With these values we can evaluate the temperature at the atoms using Eqn.~\ref{equation_effective_temperature}.  We find a stable offset between our model and the temperature measured with our in-vacuum thermometer as illustrated in Fig.~\ref{figure_bbr_main}d.

\begin{table}[h]
	\centering
	\caption{Emissivities of vacuum chamber components}
	\label{tab:emissivities}
	\begin{tabular}{  l  c  r }
		\hline
		Surface & Emissivity \\ \hline
		Viewports (glass) & 0.91 \cite{handbook2009fundamentals} \\ 
		Vacuum chamber (stainless steel) & 0.08 \cite{barnes1947total,wieting1979effects} \\ 
		Sapphire & 0.54 \cite{wittenberg1965total} \\
		Oven nozzle & 0.82 \cite{richmond1963measurement} \\
		
	\end{tabular}
\end{table}

\begin{table}[h]
	\centering
	\caption{Specular steel reflectivities based on \cite{birkebak1964effect}}
	\label{tab:reflectivities}
	\begin{tabular}{  l  r }
		\hline
		Component & Specular Reflectivity \\ \hline
		Polished Chamber & 0.95 \\ 
		Other Steel Surfaces & 0.1 \\ 
		
	\end{tabular}
\end{table}

\begin{table}[H]
	\centering
	\caption{Vacuum chamber generalized exchange factors}
	\label{tab:exchange_factors}
	\begin{tabular}{  l r}
		\hline
		Component & Exchange Factor   \\ \hline
		Heated sapphire window & $8.81\times10^{-4}$ \\ 
		2.75" CF viewport glass - extended flanges & $1.47\times10^{-2}$ \\ 
		2.75" CF viewport glass - direct flanges & $1.18\times10^{-1}$\\ 
		6" CF viewport glass & $5.39\times10^{-1}$ \\ 
		Glass cell & $1.08\times10^{-3}$ \\ 
		Metal chamber, tubing, and slower & $3.26\times10^{-1}$ \\ 
		Oven nozzle & $2.15\times10^{-5}$ \\ \hline 
	\end{tabular}

\end{table}

\subsection{Temperature Sensor Calibration}
\label{section_calibration}

In order to calculate the temperature at the atoms using Eqn.~\ref{equation_effective_temperature}, an array of sensors to monitor the temperature of various points on the vacuum chamber is required. This array is composed of commercially available 50 k$\Omega$ negative temperature coefficient thermistors (US Sensor PR503J2), specified by the manufacturer to be accurate to within 50 mK. Due to the relative importance of the 6" CF viewports (exchange factor of 0.539) to the thermal model, the calibration of the thermistors on these viewports is improved to a 13 mK uncertainty via comparison to a thin-film platinum resistance thermometer (TFPRT) calibrated to fixed-point realizations of the ice melting point and the gallium melting point.

\begin{figure}[h!]
	\centering
	\includegraphics[width=0.8\textwidth]{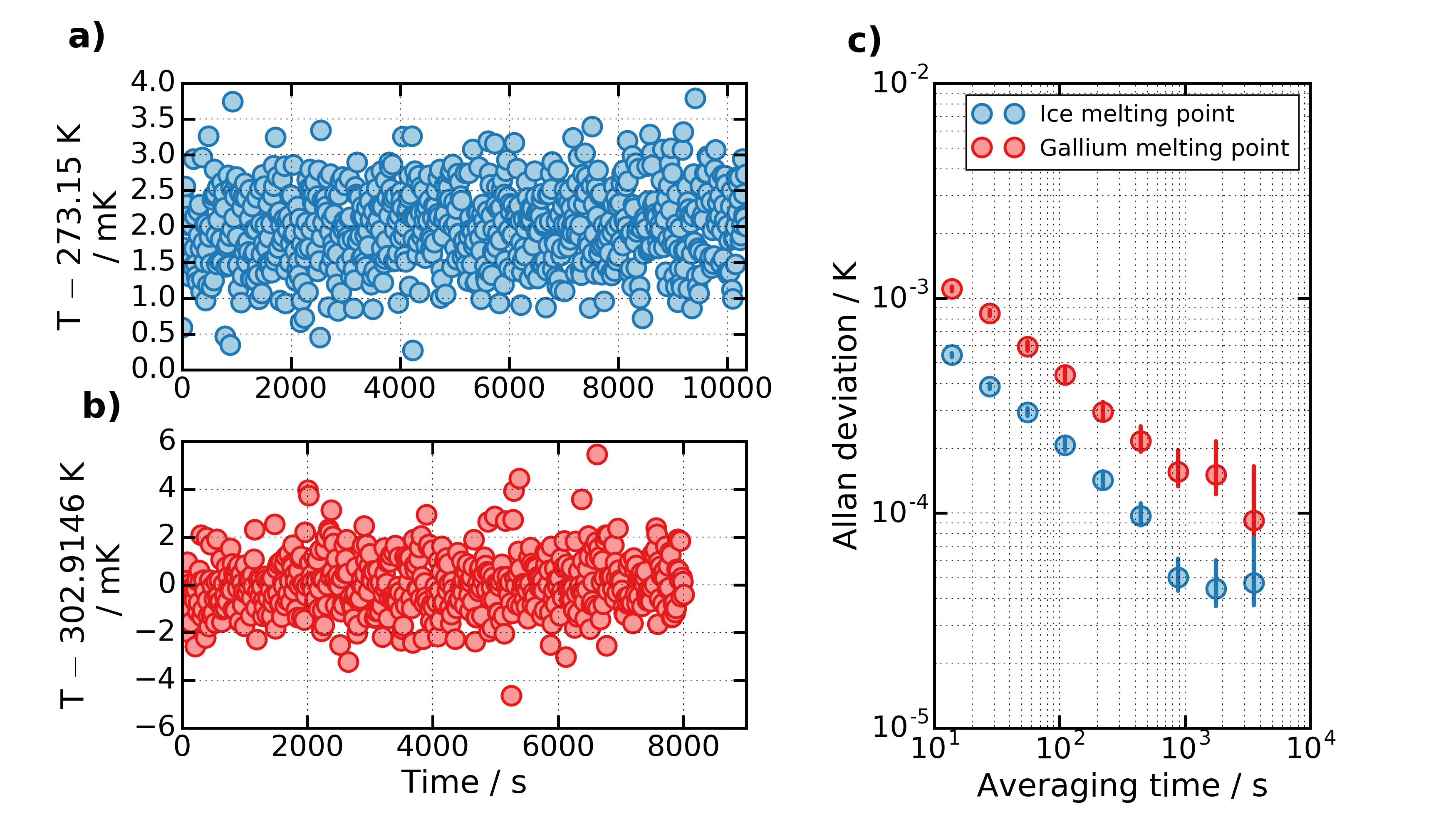}
	
	\caption{\label{fig_gallium_melt_curve}\textbf{JILA fixed-point realizations.} Data from the (a) water ice melting point and (b) gallium melting point realization. (c) Allan deviation of the gallium melt curve (red) and the ice melt curve (blue), showing the capability of averaging down to below 100 $\mu$K on each fixed-point in $10^4$ seconds or less.}
\end{figure}

Figs.~\ref{fig_gallium_melt_curve}a and \ref{fig_gallium_melt_curve}b show typical melt curves and associated Allan deviations for both the gallium ice and water ice phase transitions as measured by a TFPRT. The ITS-90 temperature scale defines these melting plateaus to be at temperatures of $29.7646^{\circ}$C (302.9146 K) and $0^{\circ}$C (273.15 K), respectively, allowing for the calibration of the TFPRT - up to systematic uncertainties - at these points. The 2 mK offset from $0^{\circ}$C in our ice point cell is a consequence of performing our calibration measurements at an elevation of $1650$ m above sea level \cite{harvey2013thermodynamic}. Dissolved air in water serves to suppress the ice point temperature at sea level, where the ice melting point corresponds to $0^{\circ}$C by definition. Upon applying this known correction, the dominant uncertainty at 22$^{\circ}$C is then the unknown non-linearity of the TFPRT between the two fixed point temperatures. After calibration of these crucial thermistors, this array of 50 k$\Omega$ thermistors is used as inputs to the thermal model. Using these inputs in conjunction with literature values for the different emissivities of the materials which compose the vacuum chamber, shown in Table~\ref{tab:emissivities}, a real-time prediction for the temperature seen by the atoms is generated.

As discussed in the main text, in order to evaluate the accuracy of the prediction generated by the ray-tracing-based thermal model, a TFPRT is installed at the position of the atoms in the vacuum chamber. This in-vacuum thermometer allows the characterization and removal of a static offset between the thermal model and the measured temperature and also serves as verification that the thermal model successfully captures all fluctuations in the temperature below the level of 1 mK. In order to achieve an accurate measure of the in-vacuum temperature from the TFPRT, the probe was hand-carried to NIST Gaithersburg for calibration in the Sensor Sciences Division facilities. There, a water bath comparison calibration with Standard Platinum Resistance Thermometers (SPRT) traceable to the ITS-90 temperature scale enabled the in-vacuum thermometer to be calibrated to an uncertainty of 1.4 mK.

\begin{figure}
	\centering
	\includegraphics[width=0.9\textwidth]{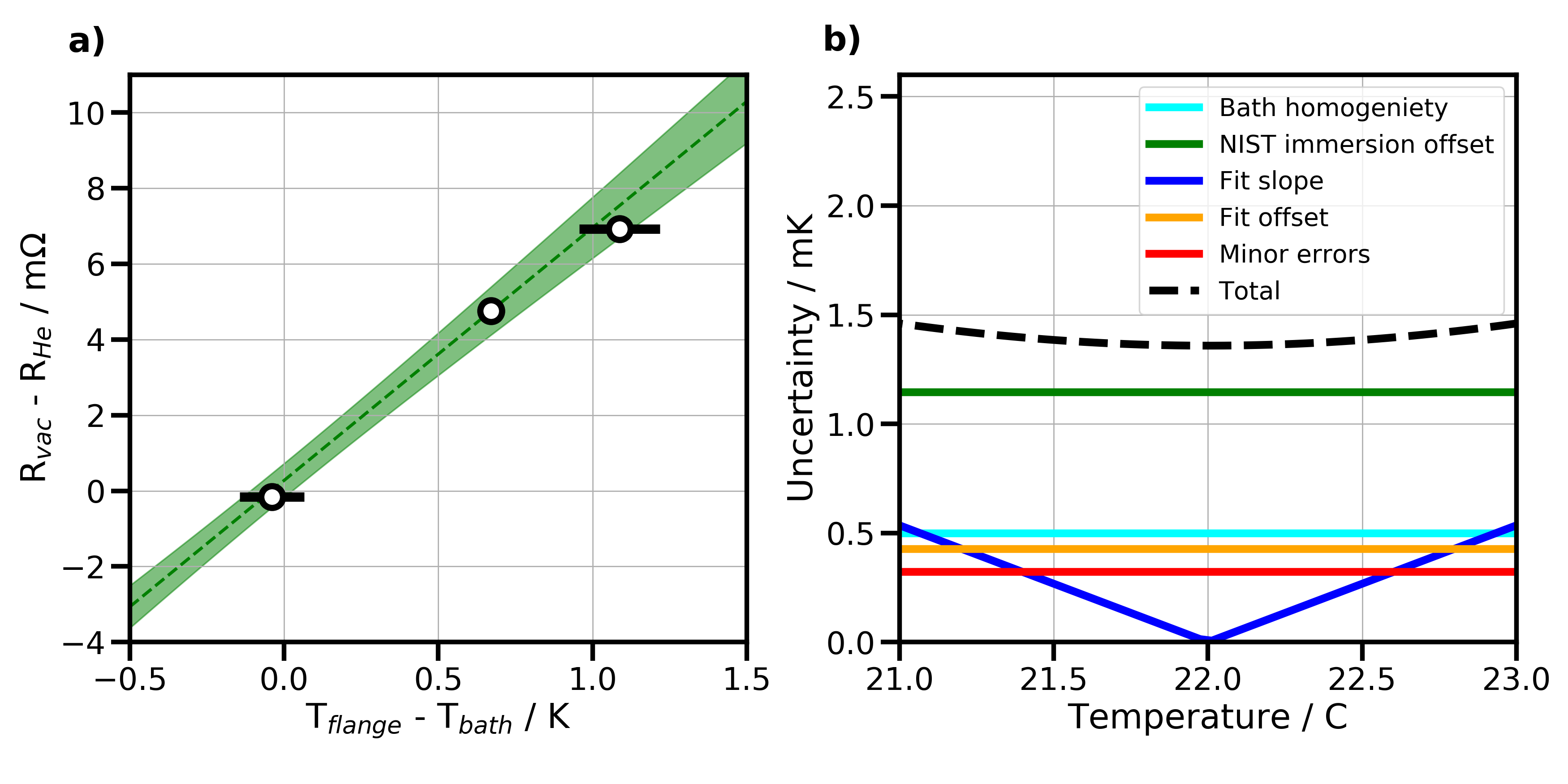}	
	\caption{\label{fig:invacuumcalib}\textbf{Calibration of the in-vacuum probe.} (a) Immersion error data in the water bath comparison. We measure the difference between the high vacuum resistance $R_{vac}$ and the He back-filled case $R_{He}$ as a function of axial gradient $T_{flange} - T_{bath}$. The data is fit to a linear function, and the fitted offset at zero axial gradient is (0.3 $\pm$ 0.4) m$\Omega$. This can be converted to temperature by using the sensitivity of the TFPRT of 2.57 Ohms/Kelvin. (b) Breakdown of uncertainties stemming from the NIST calibration. The green is the systematic error in the immersion error offset from panel (a). The cyan color is the maximum temperature gradient observed in the bath. The blue and orange curves are the fit interpolation errors from the slope and offset respectively. The red line is the quadrature sum of several minor errors relating to the calibration. The black dashed line is the quadrature sum of the errors.}
\end{figure}

Fig.~\ref{fig:invacuumcalib}b shows the different contributions to the final sensor uncertainty.
Immersion error is the largest systematic effect, arising from heat flow between the room temperature environment and the probe, producing a systematic temperature difference between the TFPRT and the bath.
To characterize this effect, a second TFPRT is mounted on the upper flange of the test chamber, which lies just above the water line of the bath. 
To evaluate immersion errors in the water bath calibration, two sets of measurements were undertaken, one under vacuum conditions ($< 3\times 10^{-6}$ Torr), and another under 30 Torr of back-filled helium. 
Figure \ref{fig:invacuumcalib}a shows the measured difference in zero power resistance for these two conditions as a function of the temperature gradient between the flange and the bath, $T_{flange} - T_{bath}$.
The immersion error is well-described by a linear function of temperature, with a fitted $T_{flange} - T_{bath} = 0$ systematic offset of (0.3 $\pm$ 0.4) m$\Omega$, which can be converted to (0.69 $\pm$ 1.1) mK. 
Since the offset is consistent with zero, we simply take the standard uncertainty of 1.1 mK as the immersion error contribution, plotted as the green line in Figure \ref{fig:invacuumcalib}b.
We also include the bath homogeniety as the largest observed gradient as indicated by the two SPRT's in the bath of $\pm$ 0.5 mK, shown as the cyan line.
The calibration coefficient uncertainties are shown as the blue and orange curves.
The red line shows the sum of all other minor errors, including SPRT calibration, bridge nonlinearity, resistance standard and bath stability~\cite{tew8046}.
We also account for the offset of 3.0 mK between ITS-90 and the definition of thermodynamic temperature, and it's corresponding uncertainty of 0.3 mK is included in the minor errors~\cite{tew8046}.

\subsection{BBR Dynamic Shift Correction}

The dynamic BBR shift is described by

\begin{equation}
\Delta \nu_{dynamic} = \nu_{dyn} \bigg ( \frac{T}{T_0} \bigg )^6.
\end{equation} 

\noindent
Care must be taken since the often reported value of $\nu_{dyn}$ contains higher order terms in T ($T^8$ and $T^{10}$) \cite{safronova2013blackbody}. The reported coefficient $\nu_{dyn}$ therefore requires corrections for temperatures deviating away from $T_0 = 300$~K. For our temperature near 22$^\circ$C, we find we must correct our dynamic BBR shift by an amount of $1.48\times10^{-18}$.

\section{Lattice Light Shift Evaluation - Model Uncertainty}
\label{section_ac_stark}

Optical lattice clocks have reached an accuracy level where higher order polarizability terms require careful consideration. Two separate approaches have dealt with this to high accuracy. At NIST, the Yb clock group \cite{BrownACstark} demonstrated that in a system where the spatial modes have certain scalings with trap depth the lattice light shifts can be well characterized by a so-called thermal model, containing effective terms that are linear ($\alpha^*$) and quadratic ($\beta^*$) in $U$:

\begin{equation}
\Delta \nu_{AC} = \alpha^* U + \beta^* U^2.
\end{equation} 

\noindent
The dipole polarizability $\alpha_{\text{E1}}$ and multipolarizability $\alpha_{\text{QM}}$ are contained in $\alpha^*$ while the hyperpolarizability $\beta$ is contained in $\beta^*$. This approach is robust as it does not require explicit knowledge of atomic coefficients or the axial and radial atomic mode numbers. A second approach is to work on a microscopic model, where for a given radial temperature and axial mode occupation a light shift can be calculated \cite{KatoriACStark}. While a microscopic model offers the prospect of accuracy below the $10^{-18}$ level, it requires significantly more input information, a potential source of systematic errors in reporting atomic coefficients. Recent work~\cite{KatoriYbACStark} has highlighted this importance by illustrating that different methods of preparing the atomic sample can result in different measurement values of $\beta$ in Yb. 

For determining the operational light shift in the SrI clock the thermal model is employed. This avoids reliance on atomic coefficients and characterizes an experiment-specific light shift model. The characterization in the main text discusses the results of the fitting of the thermal model to the data. The limitation to this analysis is that it requires addressing model error - deviations from linear and quadratic behavior in light shifts that are described by the microscopic model. To address the deviations from linear and quadratic scalings in the SrI light shifts we adopt the approach in \cite{BrownACstark}, writing the light shift as:  

\begin{equation}
    \frac{\Delta \nu_{\text{AC}}}{\nu_{\text{clock}}} = -A_n \alpha_{\text{E1}} U - B_n \alpha_{\text{QM}} U - C_n \beta U^2 
    \label{eq_microscopic_thermal}
\end{equation}

\noindent
where $A_n$, $B_n$, and $C_n$ are spatial averages given by
\begin{align}\nonumber
A_n &= \left\langle \mathbf{n} \Bigg\vert \exp\big(-\frac{2 (x^2+y^2)}{w_0^2}\big) \cos^2(kz) \Bigg\vert \mathbf{n} \right\rangle \\ \nonumber
B_n &= \left\langle \mathbf{n} \Bigg\vert \exp\big(-\frac{2 (x^2+y^2)}{w_0^2}\big) \sin^2(kz) \Bigg\vert \mathbf{n} \right\rangle \\ \nonumber
C_n &= \left\langle \mathbf{n} \Bigg\vert \exp\big(-\frac{4 (x^2+y^2)}{w_0^2}\big) \cos^4(kz) \Bigg\vert \mathbf{n} \right\rangle.
\end{align}

\noindent
The $\frac{1}{e^2}$ beam radius is given by $w_0$, the lattice wavenumber by $k$, and $ \vert \mathbf{n} \rangle = \vert n_x,n_y,n_z  \rangle$ where $n_i$ is spatial coordinate $i$'s mode number. To address gravitational sag resulting from the $\theta = 19^\circ$ tilt with respect to vertical in the lattice, the component of gravity ($g$) along $\hat{x}$ is included by substituting $x \rightarrow x-mg\sin(\theta)/\omega_r^2$ where $\omega_r$ is the radial trapping frequency.

The gravitational tilt lifts the radial degeneracy so $A_n$, $B_n$, and $C_n$ are broken into orthogonal bases such that $A_n = A_{nx} A_{ny} A_{nz}$, $B_n = B_{nx} B_{ny} B_{nz}$, and $C_n = C_{nx} C_{ny} C_{nz}$, expanding each to fourth order. The system is then described by following series of equations:

\begin{subequations}
	\begin{align}
		\gamma = \frac{g^2 \sin^2(\theta)}{w_0^2 \omega_r^4} \\
		A_{nx} = 1-\frac{\sqrt{2}}{w_0k \sqrt{U}}(n_x+1/2)(1-6\gamma)+\frac{3}{2 k^2 w_0^2 U}(n_x^2+n_x+1/2)-2\gamma+2\gamma^2 \\
		B_{nx} = A_{nx} \\
		C_{nx} =  1-\frac{2\sqrt{2}}{w_0k \sqrt{U}}(n_x+1/2)(1-6\gamma)+\frac{3}{k^2 w_0^2 U}(n_x^2+n_x+1/2)-4\gamma+4\gamma^2 \\	
		A_{ny} = 1-\frac{\sqrt{2}}{w_0k \sqrt{U}}(n_y+1/2)+\frac{3}{2 k^2 w_0^2 U}(n_y^2+n_y+1/2) \\
		B_{ny} = A_{ny} \\
		C_{ny} =  1-\frac{2\sqrt{2}}{w_0k \sqrt{U}}(n_y+1/2)+\frac{3}{k^2 w_0^2 U}(n_y^2+n_y+1/2)\\
		A_{nz} = 1-\frac{(n_z+1/2)}{\sqrt{U}} + \frac{(n_z^2+n_z+1/2)}{2U} \\
		B_{nz} = \frac{(n_z+1/2)}{\sqrt{U}} - \frac{(n_z^2+n_z+1/2)}{2U} \\
		C_{nz} = 1-\frac{2(n_z+1/2)}{\sqrt{U}} + \frac{5(n_z^2+n_z+1/2)}{2U}.
	\end{align}
\end{subequations}

\noindent
In the above equations $U$ is the trap depth in units of the photon recoil energy $E_r$.

\begin{figure}[H]
	\centering
	\includegraphics[width=0.7\textwidth]{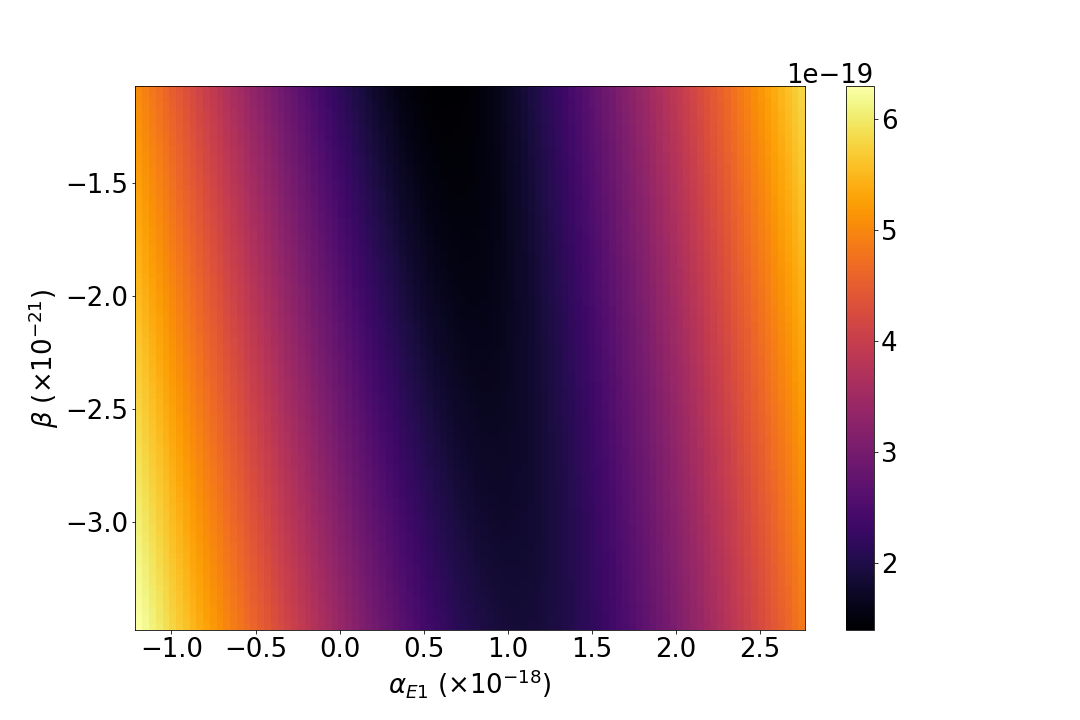}	
	\caption{\label{fig:std}\textbf{Model Error} Colormap showing the standard deviation between simulated AC Stark data (Eqn. \ref{eq_microscopic_thermal}) and a simple linear and quadratic fit to the data (Eqn. \ref{eq:thermalmodel}). The average standard deviation is 3.3$\times10^{-19}$ which we take to be our model error.}
\end{figure}

To quantify the model error, Eqn. \ref{eq_microscopic_thermal} is utilized with input data from measured $n_i$ values and $\alpha_{QM}$ from \cite{KatoriACStark}. The axial and radial mode occupation numbers are derived from carrier sideband asymmetry measurements, radial trapping frequencies, and transverse clock spectroscopy \cite{blatt2009rabi}. The final two inputs for the model, $\alpha_{E1}$ and $\beta$, are evaluated at each point in a two dimensional space of input parameters spanned by plausible values of $\alpha_{E1}$ and $\beta$. Limits on these parameters are chosen to encompass the full range of $\alpha_{E1}$ values from the four curves in Fig. \ref{fig:DensityACstark}b. Similarly, the range for $\beta$ is chosen to encompass the weighted mean of values from published literature \cite{nicholson2015systematic,SYRTEhyperpol,KatoriACStark} to one which is consistent with what is observed in Fig. \ref{fig:DensityACstark}b. For each pair of $\alpha_{E1}$ and $\beta$ in this two dimensional space of values, Eqn. \ref{eq:thermalmodel} is used to simulate lattice light shifts over the range of trap depths used in the experimental data. The thermal model from Eqn. \ref{eq:thermalmodel} is then fit to this simulated data. For each set of simulated and fitted curves, the mean and standard deviation of the difference between the two models is calculated. All mean differences are below $1\times10^{-19}$, ensuring that the standard deviation of the differences is not overlooking a constant offset between curves. The resulting average of all standard deviation values shown in Fig. \ref{fig:std} is $3.3 \times10^{-19}$. This number is thus interpreted as the model error. We note that use of the theory value for $\alpha_{\text{QM}}$ \cite{MariannaPol} provides a lower estimate of model uncertainty.

We additionally investigate the effect of a running wave in our system arising from reflectivity on the viewports. We find that the only nonlinear and nonquadratic additional term provides a negligible $2\times10^{-20}$ level effect.

\bibliographystyle{ieeetr}

\end{document}